\documentclass[12pt]{amsart}
\usepackage[left=80pt,right=80pt,top=0.75in]{geometry}
\usepackage{xcolor}
\usepackage{breakcites}
\usepackage[breaklinks=true]{hyperref}
\usepackage{amsaddr}
\usepackage{graphicx}
\usepackage{amssymb}
\usepackage{epstopdf}
\usepackage{comment}
\usepackage{csquotes}
\usepackage{fancyhdr}

\title[]{Corruptomics}

\author[]{José R. Nicolás-Carlock$^{1,*}$ \& Issa Luna-Pla$^{1}$}
\address{$^1$Institute of Legal Research, National Autonomous University of Mexico, Mexico}
\email{*jnicolas@unam.mx}

\begin{document}

\textsl{\small This is a preprint of the following chapter: Nicolás-Carlock J.R. \& Luna-Pla I., Corruptomics, published in Corruption Networks, edited by Granados O.M., Nicolás-Carlock J.R., 2021, Springer, reproduced with permission of Springer. The
final authenticated version is available online at \url{https://doi.org/10.1007/978-3-030-81484-7_9}}
\bigskip
\bigskip

\begin{abstract}
Corruption studies must evolve to match the complexity of the modern world. Here, we present three main problems in corruption analysis that need to be address: the complexity of the corruption phenomenon itself and its context, the complexity of the analytical description, and the complexity of the perspectives that different disciplines bring to the table. In this regard, we argue that the interdisciplinary framework of complex systems and network science represent a promising analytical approach to move forward in this endeavor. Furthermore, current research efforts in this direction indicate the dawn of a new interdisciplinary discipline for corruption studies in the 21st century. 
\end{abstract}   


\maketitle


\begin{displayquote}
\textsl{\small ``Injustice anywhere is a threat to justice everywhere. We are caught in an inescapable network of mutuality, tied in a single garment of destiny. Whatever affects one directly, affects all indirectly.'' - Martin Luther King Jr.}
\end{displayquote}
\bigskip


Corruption is one of the most prominent global policy challenges of the 21st century. During the last few decades, corruption not only has been extensively addressed in the public policy arena but also, it has become a very active academic research field \cite{rose2016corruption,mungiu2020research}. As an academic subject, corruption is mostly regarded as a fundamental societal problem that researchers from diverse disciplinary traditions aim to address along four main interdependent axes: conceptualizations and definitions, measuring methods and techniques, modelling of causes and consequences, and control or tackling strategies. However, although relevant advances have been made, the challenge to design theoretical and technical frameworks that are able to handle the complexity of this phenomenon remains highly contested \cite{hough2017analysing}. This is due to three main problems in corruption analysis: (I) the complexity of the nature of the phenomenon itself and its context, (II) the complexity of the analytical description, and (III) the complexity of the different perspectives that each discipline brings to the table.



\textsl{I. The complexity of the nature of the phenomenon itself and its context.} Human beings are complex and corruption is inherently hard to tackle due to the complex nature of human behavior. As any other human activity, corruption occurs within the intricate structure and dynamics of the social, economic and political systems of society. As such, corrupt behavior manifests as a non-separable activity that tends to be hidden and interwoven within the multiple activities that could be deem as non-corrupt for a given setting, from the micro group dynamics that take place within the structure of government institutions, small or big corporations, and civil life, to the macro interactions that take place among them. In addition, the systems that characterize our complex societies are not independent of each other but constitute a system of systems whose structure and dynamics are always evolving in response to changes in the corresponding social and regulation context. Notably, global dynamics have radically changed in the last few years. The flow of people, materials, and information have remarkably increased the levels of interactions at different spatial and administrative scales and consequently, the interdependencies among social, economic and political systems have become stronger. Our heterogeneous world is more connected than ever. This not only represents great cooperation opportunities for development but also highly systemic threats at regional and global scales, what happens in one place or sector can have effects on other seemingly unrelated ones, putting the analysis of corruption on a whole another level \cite{helbing2013globally, balsa2020deglobalization}. 

\textsl{II. The complexity of the analytical description.} The analysis of corruption is complex and has evolved over a long period of time. Nowadays, this analysis is done under an international consensus that has put individual behavior and indiscretions at the center of modern corruption thinking \cite{hough2017analysing}. Values, norms and ideas, that still play an important role in our understanding of what is acceptable or not in society, vary from place to place, and from time to time, making an objective understanding of moral and ethical issues a great challenge \cite{capraro2018grand}. In an effort to be more objective, the analysis of corruption has opted for a more pragmatic and empirical approach. 

On the conceptual dimension, contemporary thinking is dominated by four main approaches: legal definitions, the ``abuse of entrusted power'' criterion, economic or business-oriented, and the so-called ``legal'' corruption. Among these, the first one has become the most popular since its adoption from Transparency International, the Organization for Economic Cooperation and Development, the World Bank and the International Monetary Fund. However, the subjective elements embedded in this definition, such as what constitutes ``abuse'' and its bias towards the public sector, make it object of continuous critic and debate. On the measuring front, aggregated indices rise as the most popular as they have provided an overall and broad picture of global corruption. These still face criticism due to changes in their methodology or the perception/experience-based analysis since that makes the interpretation or comparison of results difficult. This has lead to the creation of other promising indicators that address more specific matters, as well as more sophisticated approaches based on proxies to corruption that have the capacity to describe macro features from micro data. On the modelling challenge, the consensus is low on what are the general causes of corruption. The different models based on structural forces, rational agents, principals or discretionary criteria provide insight for some settings but not all, therefore, these too are not free from debate due to the emergence of interesting collective dilemmas and the clash of social realities across the globe. On the control arena, the strategies to tackle corruption are not universal and, in direct relation with the causes of the problem at hand, strategies depend on the specific context. National regulations and international instruments aim to solve this problem but their results are not clear and there's no catalogue of solutions that work for all places at all times \cite{hough2017analysing}. 


\textsl{III. The complexity of different disciplinary perspectives.} As multifaceted as the phenomenon is, corruption analysis has been greatly enriched by the insights of researchers from different disciplines and schools of thought. To this day, these are mostly from political science, economy, sociology, anthropology, or law. In the effort to come up with a general interdisciplinary corruption theory, these researchers have to deal not only with the complex nature of phenomenon itself, the inherent heterogeneity and complexity of the systems where it takes place, or the subtleties of the four dimensions that comprise modern analysis, but also, with the clash of different ideas and perspectives about social reality that each discipline brings to the table \cite{de2010good}. This is a complex scenario where consensus on fundamental aspects is hard to achieve and, therefore, a general interdisciplinary theory of corruption has remained elusive \cite{hough2017analysing}. 


The challenge is great and in order to move forward, it's imperative to adopt new ideas and perspectives that enable us to handle and embrace the complexity of our world instead of avoid it. Here, complexity or complex systems science represent a promising approach in this endeavour:

\begin{itemize}

\item First, we must understand that the complexity of society is nothing but the result of our collective doing as we become and create the systems that shape our social, economic and political environments. In other words, we not only are but also create the complex networks in which corruption takes place. Therefore, our connections and interactions at the different scales, sectors and regions keep valuable information that can be tapped, modelled and studied in order to understand the principles that govern such systems and give us the opportunity to develop strategies and interventions tailored to the structure and dynamics of the problem at hand. In that regard, complexity science is an interdisciplinary approach to the study of collective phenomena in natural, social and technical systems that has been successful in the analysis of the structure and dynamics of such systems in terms of the relationships among their parts and their environment \cite{bar2004making,Barabasi2016}. The main idea in complex systems is that a collection of interacting components behaves in way not predicted by the components in isolation or disconnected. Interactions and dependencies matter more than the nature of the parts. Therein, the collection of interacting parts is best understood as a whole, rather than disconnected. As such, when corruption behavior manifest as a non-separable activity that tends to be hidden and intertwined among the multiple activities that occur within the structure and dynamics of social, economic, political and technical systems, then, activities that could be deemed as corrupt are best understood systemically, this is, from the collective behavior and features of connected individuals or organizations acting as a whole, rather than from their particular characteristics in isolation.

\item Second, given the multifaceted nature of corruption, the importance of a comprehensive and general analytical framework -- not necessarily universal but that unifies the different disciplinary perspectives -- and that encompasses proper conceptualizations and definitions of corrupt practice cannot be overstated, given that definitions determine what gets modelled and what researchers look for in data, in such a way that unsuited definitions can cause misleading measurements, mistaken interpretations of causes and consequences, and ultimately inappropriate policy suggestions. Achieving a unified framework seems daunting, but here again, complexity science presents itself as an interesting and relevant example on this matter, as a discipline that on its own is dealing with a similar challenge \cite{ladyman2020complex}. For that, let us recall that complexity science is an interdisciplinary approach to the study of natural, social and technical systems that has created an still-evolving analytical framework by drawing concepts and tools from disciplines such as physics, chemistry, biology, ecology, sociology, mathematics and computer science \cite{mitchell2009complexity}. In this way, complex systems are often studied in terms of networks, self-organization, evolution, non-linearity, scaling, and emergence. Remarkably, this framework has been applied to different systems and settings leading to relevant insights into crime, terrorism, war, disease spreading, financial markets, democracy and other social subjects \cite{helbing2015saving,caldarelli2020perspective,eliassi2020science,Kertesz2020}. In the case of corruption, the concepts and tools of complexity and networks have been applied to tackle specific matters, such as the conceptualization of corruption as a networked phenomenon \cite{slingerland2018network}, measuring and modelling of political corruption \cite{Ribeiro2018,Colliri2019,solimine2020political}, corruption in public procurement and corruption scandals \cite{fieruascu2017networked, Luna-Pla2020,Wachs2020,fazekas2020corruption}, and ways to identify, counteract and control phenomena such as cartel formation, money laundering and tax evasion by means of data and artificial intelligence \cite{wachs2019network,garcia2020ai,zumaya2021identifying}.

\item Lastly, anti-corruption is not an endeavor of isolated and disconnected individuals. The efforts made in the public policy arena have shown that groups of people, private institutions, and governments must learn to cooperate and work purposely in order for any strategy to be effective and successful. In this world, we are not only connected at multiple scales but also we are interdependent members of this highly complex system known as society: whatever affects one directly, affects all indirectly. Complex corruption networks are tackled with complex anti-corruption networks, and corruption studies and anti-corruption strategies must evolve to match the complexity of our reality. 

\end{itemize}


Modern technological advances by themselves are not sufficient to suggest that we have the upper-hand in the fight against corruption since, in the same way those emerging technologies allow for greater transparency, cooperation and development, they could also allow for more sophisticated corruption mechanisms and challenges. It comes down to us to make the best out of the modern tools and achievements obtained so far in the long history of corruption analysis and move forward. The research done so far to address corruption, such as the one presented all along the chapters of this book, constitute an effort in this direction, as well as example of the scope, possibilities, and potential of a new approach to corruption analysis that is open to perspectives, methods and disciplines that go beyond the traditional schools of thought in order to give place to something different and hopefully useful. 

We might not be any closer to a corruption free era, but we are positive that we are at the dawn of a new paradigm in corruption and anti-corruption studies that could take us closer to that goal. Let us consider this an opportunity to create a new discipline that embraces the complexities of the phenomenon, that takes into account the structure and dynamics of the networks that shape our societies, that takes advantage of technology, data and empirical evidence, that pulls insight from the full spectrum of the prism of interdisciplinary science, and finally, that dares to take corruption studies closer to an exact science. Let us consider and know henceforth this new discipline as \textsl{corruptomics}: corruption analysis for the 21st century.

\bibliographystyle{ieeetr}
\bibliography{References}

\bigskip
\bigskip
\textsl{\small This is a preprint of the following chapter: Nicolás-Carlock J.R. \& Luna-Pla I., Corruptomics, published in Corruption Networks, edited by Granados O.M., Nicolás-Carlock J.R., 2021, Springer, reproduced with permission of Springer. The
final authenticated version is available online at \url{https://doi.org/10.1007/978-3-030-81484-7_9}}

\end{document}